\documentclass[a4paper,12pt]{article}

\usepackage[cp866]{inputenc}
\usepackage[T2A,T1]{fontenc}

\usepackage{graphicx,epsfig}

\usepackage{amsmath,amssymb,amsfonts,amssymb}

\usepackage{authblk}
\usepackage{appendix}

\makeatletter \@addtoreset{equation}{section} \makeatother

\begin{document}

\title{Systemic Interbank Network Risks in Russia}
\author[1,2,3 \footnote{Based on the talk given by A. Leonidov at the workshop "Random Graphs and Their Applications"\;, Moscow, October 2013}]{A.V. Leonidov}
\author[2,4 \footnote{\indent The material presented does not necessarily reflect the position of the Bank of Russia on the issues under discussion. }]{E.L. Rumyantsev }
\affil[1]{Theoretical Physics Department, P.N. Lebedev Physical Institute, Moscow}
\affil[2]{Chair of Discrete Mathematics, Moscow Institute of Physics and Technology}
\affil[3]{Laboratory of Social Analysis, Russian Endowment for Science and Education, Moscow}
\affil[4]{Department of Financial Stability, Bank of Russia, Moscow}

\date{}

\maketitle

\renewcommand{\abstractname}{Abstract}

\begin{abstract}
Modelling of contagion in interbank networks is discussed. A model taking into account bow-tie structure and dissasortativity of interbank networks is developed. The model is shown to provide a good quantitative description of the Russian interbank market. Detailed arguments favoring the non-percolative nature of contagion-related risks in the Russian interbank market are given.
\end{abstract}

\newpage

\section{Introduction}

Quantitative analysis of systemic risks in financial networks presents one of the most important applications of network theory related ideas. These studies belong to a wide strand of literature devoted to analysis of cascading failures in complex networks, see e.g. \cite{Borge_2013,Pastor_2014}. One of the important topics is here developing network-based mathematical models of contagion propagation in interbank markets \cite{Kapadia_2010,Lo_2011,Caccioli_2012,Leonidov_2014}. Constructing relevant mathematical models of default propagation is nontrivial due to the necessity of reproducing such observed features of these networks as their scale-free nature and significant dissasortativity and clustering \cite{Santos_2010,Leonidov_2012,Leonidov_2013} and bow-tie structure \cite{Leonidov_2012,Leonidov_2013,Leonidov_2014}. In the literature one can find methods of taking into account dissasortativity \cite{Newman_2002_b,Boguna_2005,Gleeson_2008} and clustering \cite{Hackett_2011,Melnik_2011}, but these and similar considerations have to be transplanted into developing mathematical models of interbank loan networks allowing to reproduce their main features.

\section{Interbank network and contagion}

\begin{figure}[h!]
\centering
\includegraphics[height=0.45\textheight]{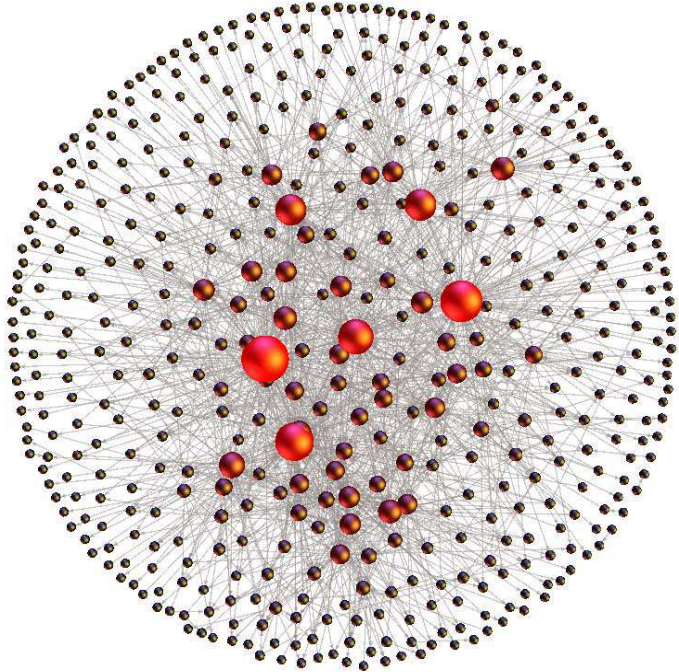}
\caption{Russian interbank network}
\label{IBRF}
\end{figure}

In what follows we characterize interbank credit market in terms of a weighted oriented graph characterized by the weighted adjacency matrix $W=\{ w_{ij} \geq 0 \}$ where link variables $w_{ij}>0$ correspond to netted obligations of the bank $i$ towards the bank $j$ on the daily basis. A directed link $i \to j$ corresponds to a credit to $i$ provided by $j$. For a given node outgoing links correspond therefore to its obligations towards neighboring nodes and incoming ones to claims of the node under consideration towards neighboring nodes so that default contagion propagates through the outgoing links. Out considerations are based on the data on Russian interbank market (see Fig.~\ref{IBRF})
from January 11, 2011 till December 30, 2013 and contain information on interbank loans to residents for 185 banks.

\section{Gai - Kapadia contagion model}

\begin{figure}[h]
\begin{minipage}[h]{\linewidth}
\centering
\includegraphics[width=\textwidth]{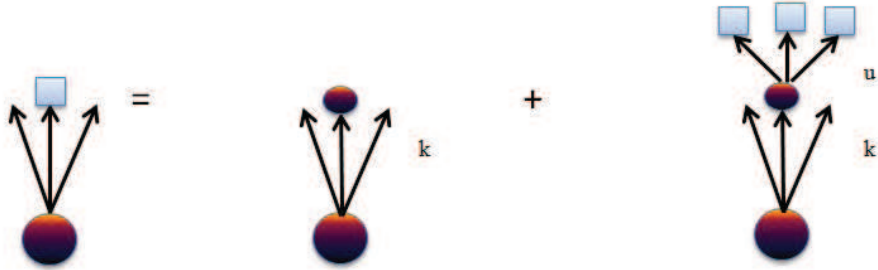}
\end{minipage}
\caption{Default propagation mechanism}
\label{contagion}
\end{figure}

Let us turn to describing the mechanics of default propagation in more details and consider a network node $i$ having $k$ incoming links with weights $\{ w_{ki} \}$, $j=1, \cdots, k$ corresponding to obligations of these $j$ banks with respect to the bank $i$. Let us now assume\footnote{Here and in what follows we consider only the simplest case in which we have only one originally defaulting node.} that one of these $j$ banks $j^*$ defaults. The default propagates (contagion takes place) if this causes default of the node $i$, see Fig.~\ref{contagion}. In the simplest setting this happens when the loss of $w_{j^* i}$ destroys the institutionally required balance between $j$'s actives $A_i$ and obligations $L_i$. The model of \cite{Kapadia_2010} assumes that for all     nodes the sum of all node's interbank assets $A_i^{\rm IB}$, $A_i^{\rm IB} \equiv \sum_{j=1}^k w_{jk}$, is evenly distributed over $k$ incoming links. A default of any of the neighboring nodes causes contagion if it makes the capital buffer $K_i \equiv A_i-L_i$ negative\footnote{Here it is assumed that the institutional requirement on the capital buffer is $K_i \geq 0$}:
\begin{equation}
K_i \equiv A_i-L_i < \dfrac{A_i^{\rm IB}}{k}
\end{equation}
A tractable analytical model of default propagation developed in \cite{Kapadia_2010} is based on a probabilistic description of contagion by introducing a probability $v_k$ that the bank $i$ is vulnerable, i.e. that it defaults because of defaulting of one of its counterparties:
\begin{equation}
v_k = {\rm Prob} \left[ K_i  < \dfrac{A_i^{\rm IB}}{k} \right]
\end{equation}
A process of contagion is in these terms that of formation of a cluster of vulnerable nodes formed around the initially defaulting one, see Fig.~\ref{contagion}.
Assuming the absence of correlations between the adjacent nodes the full probabilistic description of a problem can then given in terms the following generating function:
\begin{equation}
{\cal G} (x,y) = \sum_{j,k} v_j p_{jk} x^j y^k,
\end{equation}
where $p_{jk}$ is a probability of having $j$ incoming and $k$ outgoing links.

Introducing the generating function $G_0(y)$ and $G_1(y)$ for the out-degrees of the vulnerable bank and its vulnerable neighbor respectively
\begin{equation}
G_0(y)  =  \sum_{j,k} v_j \cdot p_{jk} \cdot y^k, \;\;\;\;\;\;G_1(y)  =  \dfrac{\sum_{j,k} v_j \cdot j \cdot p_{jk} \cdot y^k}{\sum_{j,k} j \cdot p_{jk}}
\end{equation}
we get a standard set of equations for the size of vulnerable cluster reached by following and arbitrary link leading from the initial vulnerable node $H_1(y)$ and the total size of the vulnerable cluster $H_0(y)$:
\begin{eqnarray}
H_0 (y)& = & 1-G_0(1)+y G_0[H_1(y)]^k \nonumber \\
H_1(y) & = & 1- G_1(1) + y G_1[H_1(y)]
\end{eqnarray}
and, consequently, the following equation for the average size $S$ of a default cluster formed by following the outgoing links joining vulnerable nodes:
\begin{equation}\label{S_GK}
S = G_0(1) + \dfrac{G'_0(1) G_1(1)}{1-G'_1(1)}
\end{equation}
In Ref.~\cite{Kapadia_2010} systemic risk was defined as an appearance of the giant vulnerable cluster at the point $G'_1(1)=1$. Let us note that numerical simulations in \cite{Kapadia_2010} were performed using the Poissonian distribution for the number of outgoing links.

The analysis of \cite{Kapadia_2010} was further developed in \cite{Caccioli_2012} where a dependence of percolation threshold on replacing poisson degree distribution by a scale-free one and on degree-degree correlations was studied.

\section{Contagion model with bow-tie structure and dissasortativity}

In this section we discuss a model of contagion propagation in interbank markets explicitly taking into account the bow-tie topology of the corresponding network, its dissasortativity and the tree-like nature of default clusters\footnote{This observation is similar to the one made in \cite{Goel_2013}. We are grateful to C. Borgs for this reference.} \cite{Leonidov_2012,Leonidov_2013,Leonidov_2014}. The consideration in \cite{Leonidov_2012,Leonidov_2013,Leonidov_2014} uses the daily data on the Russian interbank market. The model accounts for detailed probabilistic patterns existing between adjacent nodes characterized by conditional probabilities of default propagation $v(u,t|k,l)$ and link multiplicity $P(u,t|k,l)$, where $u$ and $t$ are the number of incoming and outgoing links for the lender while $k$ and $l$ are the numbers of incoming and outgoing links for the borrower. The conditional probabilities $v(u,t|k,l)$ and $P(u,t|k,l)$ depend on the position of the corresponding nodes in the bow-tie structure so that the formalism includes conditional probabilities $v^{\rm IO \to IO}(u,t|k,l)$ and $P^{\rm IO \to IO}(u,t|k,l)$ if both nodes belong to the In-Out component, etc.

\subsection{Existence of giant cluster}

\begin{figure}[h!]
\begin{minipage}[h]{\linewidth}
\centering
\includegraphics[width=\textwidth]{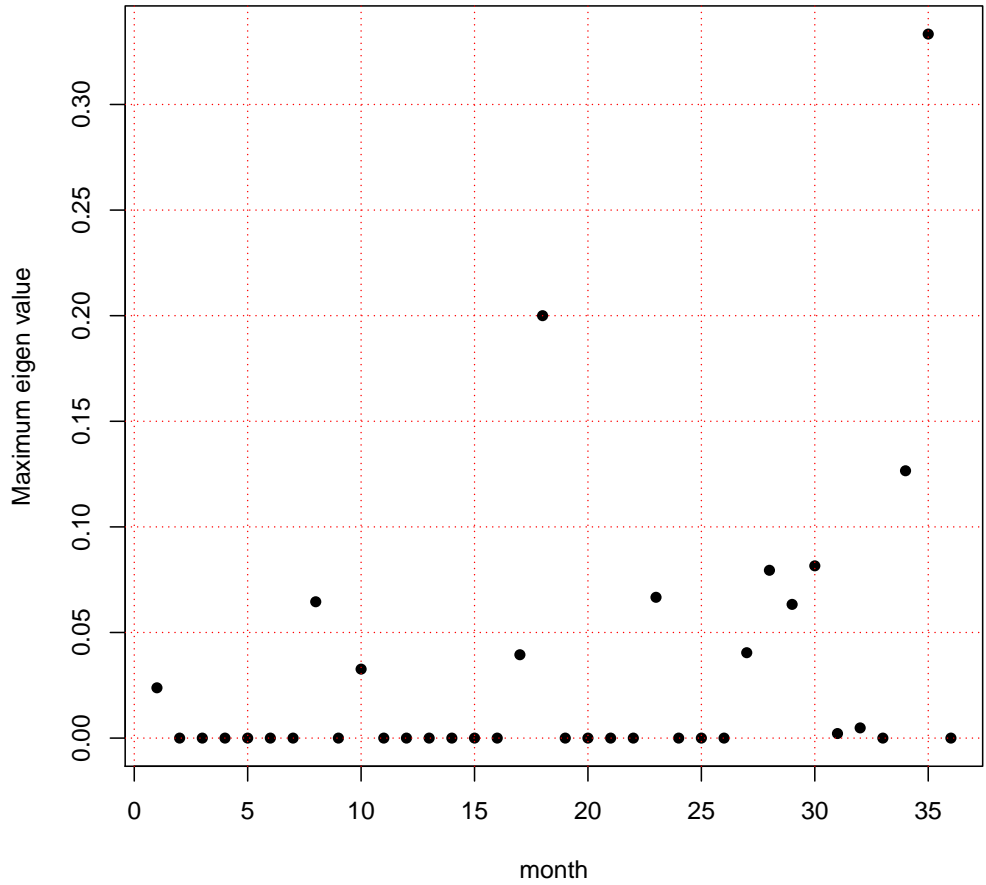}
\end{minipage}
\label{FEigenValue}
\caption{Dynamics of the maximal eigenvalue of $A$. The data was sampled with a frequency of one month.}
\end{figure}

We have already mentioned a theoretically appealing definition of systemic risk as of the appearance of the giant cluster \cite{Kapadia_2010}.The practical relevance of this definition does however depend on whether formation of such a cluster is possible in real interbank networks. To answer this question one has to convert information on interbank loans and bank balance sheets into conditional probability distributions $v(u,t|k,l)$ and $P(u,t|k,l)$ thus specifying the structure of clusters of vulnerable nodes. For tree-like oriented graphs the condition of existence of a giant out-cluster can be formulated in terms of a condition $\lambda_{\max}>1$ on the maximal eigenvalue of the matrix
\begin{equation}\label{matrixa}
A_{(k,l)(u,t)}=uP(u,t|k,l)v(u,t|k,l),
\end{equation}
see e.g. \cite{Boccaletti_2006,Boguna_2005,Gleeson_2011}\footnote{In actual calculations it is convenient to change notations so that with each pair $(k,l)$ and $(u,t)$ one associates a natural number. A discussion of the origin of the criterion $\lambda_{\max}>1$ can be found in \cite{Gleeson_2008}}. The details can be found in the Appendix. The dynamics of $\lambda_{\max}$ for the Russian interbank network is shown in Fig.~\ref{FEigenValue}.
We see that the maximal observed value of $\lambda_{\max}$ is 0.35, so according to this criterion in hte Russian interbank market the systemic risk as defined in \cite{Kapadia_2010} is absent.

\subsection{Mean default cluster size}

\begin{figure}[h]
\begin{minipage}[h]{0.95\linewidth}
\centering
\includegraphics[width=\textwidth]{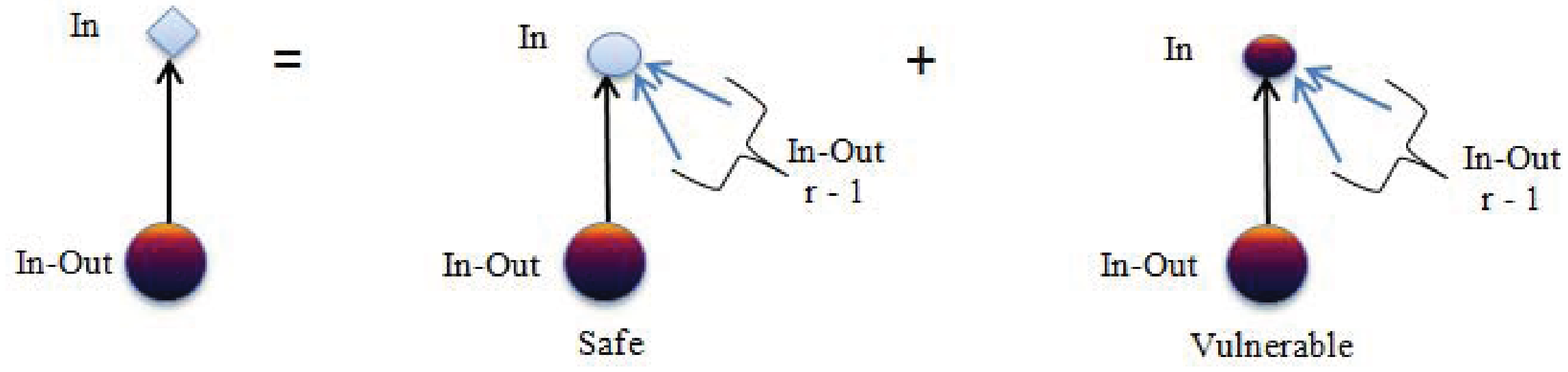} a) \\
\end{minipage}

\vfill

\begin{minipage}[h]{0.95\linewidth}
\centering
\includegraphics[width=\textwidth]{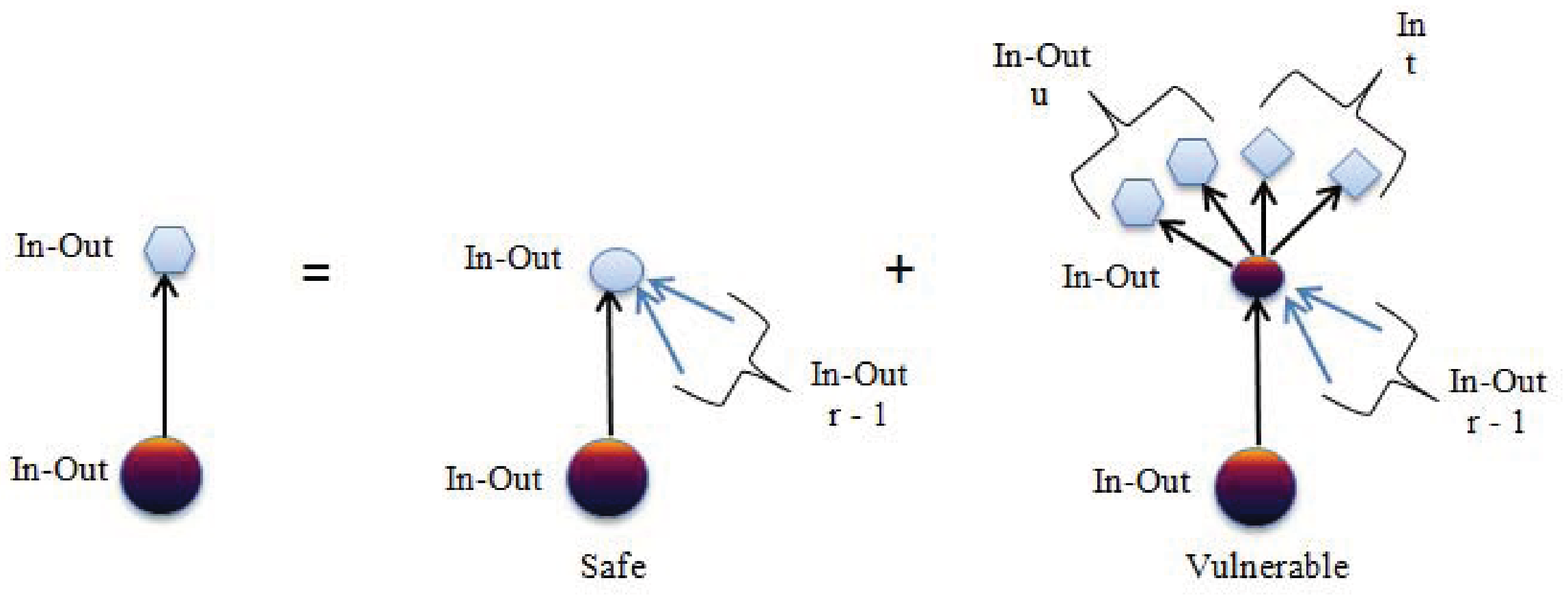} b) \\
\end{minipage}
\caption{Default spreading mechanism}
\label{FDefSpr}
\end{figure}

The mathematical model we use to describe systemic risks on the Russian interbank market developed in \cite{Leonidov_2014} generalizes the approach of \cite{Kapadia_2010} by explicitly taking into account the bow-tie structure of the network under consideration and its dissasortaivity. This means, in particular, that one has to consider two separate mechanisms of contagion propagation, those from the In-Out component to the In one, see Fig.~\ref{FDefSpr} a, and within the In-Out component, see Fig.~\ref{FDefSpr} b.

The corresponding generating functions read
\begin{eqnarray}
N_{k,l}(y) & = & \sum\limits_{r}^{\infty} P^{\rm IO \to In}(r|k,l)\left(1-v^{\rm IO \to In}(r|k,l)+y\;v^{\rm IO \to In}(r|k,l) \right) \\
M_{k,l}(x,y)& = & \sum\limits_{u,t,r}^{\infty} P^{\rm IO \to IO}(u,t,r|k,l)(1-v^{\rm IO \to IO}(u,t,r|k,l))+\\
&&x\sum\limits_{u,t,r}^{\infty} P^{\rm IO \to IO}(u,t,r|k,l) v^{\rm IO \to IO}(u,t,r|k,l)[M_{u,t}(x,y)]^{u}[N_{u,t}(y)]^{t} \nonumber
\end{eqnarray}

Let us consider a bank from the In-Out component with $k+l$ outgoing links, where $k$ of them lead to the In-Out- component and $l$ to In- component respectively\footnote{As discussed in \cite{Leonidov_2014}, nodes from the Out- component generate very small systemic risks so that the corresponding effects will be neglected} and take a randomly chosen edge linking the chosen node to a node in the In- component which, in addition, has $r-1$ incoming links, see Fig. \ref{FDefSpr} a. This is a simplest case where contagion goes from the In-Out- component to the In-one and stops there.

Let us introduce a generating function $F(x,y)=\sum\limits_{k,l}^{\infty} P^{\rm IO}(k,l)x^ky^l$ for the probability for a bank from the In-Out- component to have $k$ and $l$ first neighbors from the In-Out- and In- components respectively. Then $F(M,N)$ is the generation function for the number of vulnerable banks in the network. The mean size of vulnerable cluster $\langle s \rangle$ is then given by its derivative at $y=x=1$:
\begin{equation}
\langle s \rangle = F'_x (y=x=1)
\end{equation}
We have
\begin{equation}
F'(M,N)=\sum\limits_{k,l}^{\infty} P^{\rm IO}(k,l)(k M'_{k,l}+lN'_{k,l}),
\end{equation}
Straightforward calculations \cite{Leonidov_2014} lead to the following expressions for $N'$ and $M'$:
\begin{eqnarray}
N'_{k,l|x=1} & = & \sum\limits_{r}^{\infty} P^{\rm IO \to In}(r|k,l)v^{IO \to In}(r|k,l) \\
M'_{k,l}& = & \sum\limits_{u,t}^{\infty} \beta_{k,l,u,t} \gamma_{u,t}
\end{eqnarray}
where $\beta_{k,l,u,t}$ is an element $B_{(u,t),(k,l)}$ of the matrix $B=(I-A)^{-1}$ and $A$ is a $k \times l,k \times l$ matrix  size with the elements $A_{(k,l)(u,t)}=\alpha_{u,t,k,l}$, where in turn
\begin{equation}
\alpha_{u,t,k,l}=\sum\limits_{r}^{\infty} u P^{\rm IO \to IO}(u,t,r|k,l)v^{\rm IO \to IO}(u,t,r|k,l)
\end{equation}
and
\begin{eqnarray}
\gamma_{k,l} & = & \sum\limits_{u,t,r}^{\infty} P^{\rm IO \to IO}(u,t,r|k,l)v^{IO \to IO}(u,t,r|k,l) \nonumber \\
& + & \sum\limits_{u,t,r}^{\infty} P^{IO \to IO}(u,t,r|k,l)v^{\rm IO \to IO}(u,t,r|k,l) \nonumber \\
& \times & t \; \sum\limits_{r1}^{\infty} P^{IO \to In}(r_1|u,t)v^{\rm IO \to In}(r_1|u,t)
\end{eqnarray}

\begin{figure}[h!]
\centering
\includegraphics[width=0.45\linewidth]{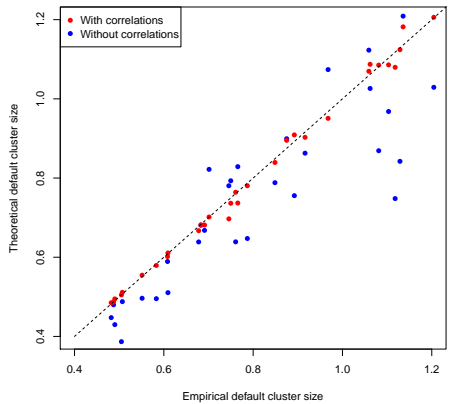}
\caption{Theoretical and empirical mean default cluster size}
\label{FDefClusterSizeCompare}
\end{figure}

Plugging in empirical conditional probability distributions  $P^{\rm IO \to In}(r|k,l)$, $v^{\rm IO \to In}(r|k,l)$, $P^{\rm IO \to IO}(u,t,r|k,l)$ and $v^{\rm IO \to IO}(u,t,r|k,l)$ calculated on the monthly basis we compute the corresponding values of $\langle s \rangle$. A comparison of the model predictions and results of stress testing are shown in Fig.\ref{FDefClusterSizeCompare}. We see a very good agreement between the model and experiment provided one takes into account correlations between the degrees of adjacent nodes captured by $P^{\rm IO \to In}(r|k,l)$ and $P^{\rm IO \to IO}(u,t,r|k,l)$ and a much poorer one when these correlations are neglected.

\section{Conclusions}

Let us formulate once again the main conclusions of the present paper:
\begin{enumerate}
\item Analysis of data on Russian interbank market shows that default contagion risks can be classified as those characteristic of non-percolative phase.
\item To build a successful mathematical model of contagion propagation the bow-tie structure of the corresponding network and its dissasortativity have to be taken into account.
\end{enumerate}

\section*{Appendix}

\appendix

Let us consider formation of giant out-cluster in an oriented graph characterized by the degree distribution, i.e. the probability of having $k$ incoming and $l$ outgoing links $p_{kl}$, and probability distribution $P(u,t|k,l)$, the conditional probability of having $u$ incoming and $t$ outgoing links for a first neighbor of the node having $k$ incoming and $l$ outgoing links.

The mean number of first neighbors that can be reached by following the outgoing links $z_1$ is conveniently computed using the corresponding generating function $F(x)$:
\begin{equation}\label{z1}
\left. z_1=\frac{dF(x)}{dx} \right \vert_{x=1}=\sum\limits_{k,l}^{\infty} kp_{kl}, \;\;\; F(x)=\sum\limits_{k=0,l=0}^{\infty}p_{kl} x^k
\end{equation}

To calculate the mean number of second neighbors let us use a construction described, e.g., in \cite{Newman_2002_a} and introduce a generating function $M_{(k,l)}(x)$ for the number of incoming and outgoing links of a vertex reached from the original one by following one of its outgoing links:
$$M_{(k,l)}(x)=x\sum\limits_{u,t}^{\infty} P(u,t|k,l)x^u$$

Let us denote the number of the $n$-th level neighbors by $z_n$ consider the total number $m_n$ of neighbors up to level $n$:
$$
m_1=z_1, \;\; m_2=z_1+z_2, \cdots, \;\; m_n=\sum_{i=1}^n z_i
$$
Then
\begin{eqnarray}
 m_2 & = & \left. \frac{d F(M(x))}{d x} \right \vert_{x=1}=\frac{\partial F}{\partial M_{(k,l)}}\frac{\partial M_{(k,l)}}{\partial x}=\sum\limits_{k,l}^{\infty} kp_{kl}+\sum\limits_{k,l}^{\infty} kp_{kl}\sum\limits_{u,t}^{\infty} uP(u,t|k,l) \nonumber \\
z_2 & = & m_2-z_1=\sum\limits_{k,l}^{\infty} kp_{kl}\sum\limits_{u,t}^{\infty} uP(u,t|k,l) \nonumber
\end{eqnarray}
Analogously
\begin{eqnarray}
m_3 & = & \left. \frac{d F(M(M,x))}{d x}\right \vert_{x=1}=\frac{\partial F}{\partial M_{(k,l)}}
\left [\frac{\partial M_{(k,l)}}{\partial M_{(u,t)}}\frac{\partial M_{(u,t)}}{\partial x}+\frac{\partial M_{(k,l)}}{\partial x} \right] \nonumber \\
& = & \sum\limits_{k,l}^{\infty} kp_{kl}+\sum\limits_{kl}^{\infty} kp_{kl}\sum\limits_{u,t}^{\infty} uP(u,t|k,l) \nonumber \\
& & +\sum\limits_{k,l}^{\infty} kp_{kl}\sum\limits_{u,t}^{\infty} uP(u,t|k,l)\sum\limits_{u_1,t_1}^{\infty} u_1P(u_1,t_1|u,t) \nonumber \\
z_3 & = & m_3-m_2=\sum\limits_{k,l}^{\infty} kp_{kl}\sum\limits_{u,t}^{\infty} uP(u,t|k,l)\sum\limits_{u_1,t_1}^{\infty} u_1P(u_1,t_1|u,t), \nonumber
\end{eqnarray}
and, generically,
$$
z_n=\sum\limits_{k,l}^{\infty} kp_{kl}\underbrace{\sum\limits_{u,t}^{\infty} uP(u,t|k,l)\cdot...\cdot\sum\limits_{u_{n-1},t_{n-1}}^{\infty} u_{n-1}P(u_{n-1},t_{n-1}|u_{n-2},t_{n-2})}_{\mbox{n summations}}.
$$
The mean number of all neighbors is thus given by
\begin{equation}
\sum\limits_{i=1}^{\infty} z_i=\sum\limits_{k,l}^{\infty} kp_{kl}[1+\sum\limits_{u,t}^{\infty} uP(u,t|k,l)+\sum\limits_{u,t}^{\infty} uP(u,t|k,l)\sum\limits_{u_1,t_1}^{\infty} u_1P(u_1,t_1|u,t)+...]
\label{eqSum1}
\end{equation}
The giant component exists if the sum (\ref{eqSum1}) diverges.

Let us now discuss the conditions for the existence of a giant component. Defining $A_{(k,l)(u,t)}=uP(u,t|k,l)$, we can rewrite (\ref{eqSum1}) as follows:
\begin{eqnarray}
\sum\limits_{i=1}^{\infty} z_i & = & \sum\limits_{k,l}^{\infty} kp_{kl}\sum\limits_{u,t}^{\infty}[I_{(k,l)(u,t)}+A_{(k,l)(u,t)}+(A^2)_{(k,l)(u,t)}+...] \nonumber \\
& = & \sum\limits_{k,l}^{\infty} kp_{kl}\sum\limits_{u,t}^{\infty}[I+A+A^2+...]_{(k,l)(u,t)}
\label{eqSum2}
\end{eqnarray}

Let us assume that $\sum\limits_{k,l}^{\infty} kp_{kl}<\infty$. For the sum (\ref{eqSum2}) to converge the operator $A$ has to be linear and bounded, $\Vert A\Vert<1$. Then there exist an operator $(I-A)^{-1}=\sum\limits_{n=0}^{\infty}A^n$ and the sum (\ref{eqSum2}) can be rewritten in the following form:
\begin{equation}
\sum\limits_{i=1}^{\infty} z_i=\sum\limits_{k,l}^{\infty} kp_{kl}\sum\limits_{u,t}^{\infty}[I-A]^{-1}_{(k,l)(u,t)}
\label{eqSum3}
\end{equation}
The condition $\Vert A\Vert<1$ leads us to a simple criterion for the absence of a giant cluster: if the maximal eigenvalue of $A$ satisfies $\lambda_{\max}<1$, there is no giant Out-component\footnote{Here we have used the spectral theorem stating that the spectral radius of $A$ is equal to its norm and the Perron-Frobenius theorem, according to which for non-negative matrix its spectral radius is equal to its maximal eigenvalue.}. Let us also note that according to Perron-Frobenius theorem the maximal eigenvalue satisfies
$$
\lambda_{\max}\leq\max_{(k,l)}\sum\limits_{u,t}^{\infty}A_{(k,l)(u,t)}
$$.

\end{document}